\documentclass[sigconf,natbib=true]{acmart}

\AtBeginDocument{%
  }

\copyrightyear{2026}
\acmYear{2026}
\acmConference[SIGIR '26]{Proceedings of the 49th International ACM SIGIR Conference on Research and Development in Information Retrieval}{July 20--24, 2026}{Melbourne, VIC, Australia}
\acmBooktitle{Proceedings of the 49th International ACM SIGIR Conference on Research and Development in Information Retrieval (SIGIR '26), July 20--24, 2026, Melbourne, VIC, Australia}
\acmDOI{10.1145/3805712.3808539}
\acmISBN{979-8-4007-2599-9/2026/07}
\usepackage{xcolor} 

\usepackage{wrapfig}
\usepackage{subcaption}
\usepackage{amsmath}
\usepackage{graphicx}
\usepackage[utf8]{inputenc} 
\usepackage[T1]{fontenc}    
\usepackage{hyperref}       
\usepackage{url}            
\usepackage{booktabs}       
\usepackage{amsfonts}       
\usepackage{nicefrac}       
\usepackage{microtype}      
\usepackage{xcolor}         
\usepackage{amsmath}
\usepackage{enumitem}
\usepackage{multirow}
\usepackage{balance}        

\title[Rethinking Semantic–Collaborative Integration: Why Alignment Is Not Enough]{Rethinking Semantic–Collaborative Integration: \\Why Alignment Is Not Enough}

\begin{document}


\author{Maolin Wang}
\email{MorinWang@foxmail.com}
\affiliation{%
  \department{Hong Kong Institute of AI for Science}
  \institution{City University of Hong Kong}
  \city{Hong Kong SAR}
  \country{China}
}

\author{Dongze Wu}
\email{dongzew@buaa.edu.cn}
\affiliation{%
  \institution{School of Artificial Intelligence, Beihang University}
  \city{Beijing}
  \country{China}
}

\author{Jianing Zhou}
\email{23371069@buaa.edu.cn}
\affiliation{%
  \institution{School of Artificial Intelligence, Beihang University}
  \city{Beijing}
  \country{China}
}

\author{Hongyu Chen}
\email{hychan2355-c@my.cityu.edu.hk}
\affiliation{%
  \department{Hong Kong Institute of AI for Science}
  \institution{City University of Hong Kong}
  \city{Hong Kong SAR}
  \country{China}
}

\author{Beining Bao}
\email{beininbao2-c@my.cityu.edu.hk}
\affiliation{%
  \department{Hong Kong Institute of AI for Science}
  \institution{City University of Hong Kong}
  \city{Hong Kong SAR}
  \country{China}
}

\author{Yu Jiang}
\email{luckyjiang@link.cuhk.edu.hk}
\affiliation{%
  \institution{Chinese University of Hong Kong}
  \city{Hong Kong SAR}
  \country{China}
}

\author{Chenbin Zhang}
\email{aleczhang13@gmail.com}
\affiliation{%
  \institution{Unaffiliated}
  \country{China}
}

\author{Chang Wang}
\email{qili9503@foxmail.com}
\affiliation{%
  \institution{Jingdong}
  \city{Beijing}
  \country{China}
}

\author{Jian Liu}
\email{liujian2024@buaa.edu.cn}
\affiliation{%
  \institution{School of Artificial Intelligence, Beihang University}
  \city{Beijing}
  \country{China}
}

\author{Lei Sha}
\authornote{Corresponding author.}
\email{shalei@buaa.edu.cn}
\affiliation{%
  \institution{School of Artificial Intelligence, Beihang University}
  \city{Beijing}
  \country{China}
}
\renewcommand{\shortauthors}{Maolin Wang et al.}

\begin{abstract}
Large language models (LLMs) have become an important semantic infrastructure for modern recommender systems. A prevailing paradigm integrates LLM-derived semantic embeddings with collaborative representations via representation alignment, implicitly assuming that the two views encode a shared latent entity and that stronger alignment yields better results. We formalize this assumption as the global low-complexity alignment hypothesis and argue that it is stronger than necessary and often structurally mismatched with real-world recommendation settings. We propose a complementary perspective in which semantic and collaborative representations are treated as partially shared yet fundamentally heterogeneous views, each containing both shared and view-specific factors. Under this shared-plus-private latent structure, enforcing global geometric alignment may distort local structure, suppress view-specific signals, and reduce informational diversity. To support this perspective, we develop complementarity-aware diagnostics that quantify overlap, unique-hit contribution, and theoretical fusion upper bounds. Empirical analyses on sparse recommendation benchmarks reveal low item-level agreement between semantic and collaborative views and substantial oracle fusion gains, indicating strong complementarity. Furthermore, controlled alignment probes show that low-capacity mappings capture only shared components and fail to recover full collaborative geometry, especially under distribution shift. These findings suggest that alignment should not be treated as the default integration principle. We advocate a shift from alignment-centric modeling to complementarity fusion-centric, complementarity-aware design, where shared factors are selectively integrated while private signals are preserved. This reframing provides a principled foundation for the next generation of LLM-enhanced recommender systems. The code for this paper is publicly available at github\footnote{\url{https://github.com/MorinW/Semantic-Collaborative-Integration}}.
\end{abstract}




\maketitle

\section{Introduction}

Large language models (LLMs) have rapidly become an important semantic infrastructure for modern recommender systems~\cite{hou2024bridging}. Recent research increasingly integrates LLM-derived representations into collaborative filtering architectures through embedding initialization, lightweight adapter tuning, contrastive alignment objectives, or explicit representation matching strategies~\cite{liu2023llmrec}. The prevailing intuition is that LLM embeddings encode rich, high-level semantic structure that is complementary to sparse user–item interaction data, and that these informative signals can be effectively incorporated into existing recommendation pipelines without incurring the substantial inference cost typically associated with fully generative models~\cite{wu2024survey}.

A dominant design pattern has therefore emerged: align semantic representations with collaborative latent spaces~\cite{wang2023collaborative}.
In most approaches, LLM embeddings are projected, transformed, or regularized so that they approximate the geometry of interaction-based embeddings~\cite{v2025contrastive}.
This strategy is motivated by an implicit hypothesis that semantic and collaborative representations capture the same underlying factors, up to a learnable transformation~\cite{liu2025enhancing}.

We formalize this hypothesis as \emph{the global low-complexity alignment hypothesis}~\cite{lin2025can}.
Under this hypothesis, semantic and collaborative spaces are presumed to be globally compatible: minimizing their discrepancy should preserve useful information and improve recommendation performance~\cite{wang2024towards}.
Stronger alignment is interpreted as better integration~\cite{zhao2024recommender}.

In this paper, we argue that this hypothesis is stronger than necessary and may not hold in general~\cite{varun2024multimodal}.
Semantic representations are shaped by language modeling objectives and content statistics, reflecting explicit attribute coherence and conceptual similarity~\cite{devlin2019bert,bge-m3}.
Collaborative representations, in contrast, are induced from interaction graphs that encode co-occurrence patterns, exposure mechanisms, popularity dynamics, and collective behavioral regularities~\cite{lightgcn}.
Although these two views overlap, they arise from different generative processes and may encode distinct, view-specific factors~\cite{dai2024bias}.
The hypothesis that they can be globally aligned overlooks the possibility that certain dimensions in one space have no meaningful correspondence in the other~\cite{radford2021learning}.

When two representations are only partially shared, enforcing global alignment may have unintended consequences~\cite{wang2015collaborative}.
First, geometric constraints imposed by alignment objectives can distort local neighborhood structure in one or both spaces~\cite{wang2023collaborative}.
Second, interaction-derived representations in sparse or long-tail regions may be dominated by noise, making strict alignment undesirable~\cite{rendle2012bpr}.
Third, if alignment collapses view-specific factors into a shared latent space, the resulting system may primarily capture the intersection of information sources rather than their full combination~\cite{varun2024multimodal}.
These concerns suggest that alignment should not be treated as a universal objective~\cite{li2024llm}.

Based on this shared-plus-private structure, enforcing global agreement is not only unnecessary but potentially harmful; we therefore advocate a shift from \textbf{alignment-centric} integration to \textbf{complementarity fusion-centric designs} that preserve private factors while selectively sharing what is truly shared~\cite{burke2002hybrid}.
In the alignment-centric view, semantic and collaborative embeddings are regarded as two noisy observations of the same latent structure, and the primary goal is to enforce agreement between them~\cite{wang2023collaborative}.
In the complementarity fusion-centric view, the two representations are treated as partially overlapping yet structurally distinct views~\cite{varun2024multimodal}.
The objective becomes preserving complementary signals while enabling controlled interaction between shared components~\cite{burke2002hybrid}.

This reframing has methodological implications~\cite{zangerle2022evaluating}.
Model design should distinguish between shared and private factors rather than enforcing global proximity. 
Complementarity fusion mechanisms can operate at the representation or score level without requiring geometric collapse into a single latent space. 
Evaluation protocols should consider not only overall ranking metrics, but also complementarity diagnostics that reveal whether two views contribute overlapping or distinct signals. 
Such diagnostics are necessary to determine whether a method genuinely expands informational coverage or merely redistributes existing evidence.

Specifically, we make the following contributions:
\begin{itemize}[leftmargin=*]
    \item We articulate \emph{the global low-complexity alignment hypothesis} as a foundational hypothesis in LLM-enhanced recommender systems and examine its limitations.
    \item We introduce a complementarity fusion-centric, complementarity-aware perspective that treats semantic and collaborative representations as partially shared but not globally alignable.
    \item We discuss the modeling and evaluation implications of this perspective, outlining a research agenda that emphasizes complementary signal preservation and selective information sharing.
\end{itemize}

\textbf{Significance.}
As LLM-derived semantics become increasingly embedded in retrieval and recommender systems, clarifying the structural relationship between semantic and collaborative views is essential. 
The proposed perspective encourages future work to explicitly reason about representation heterogeneity, shared versus private factors, and trade-offs between alignment and diversity. 
By shifting attention from enforcing agreement to managing complementarity, this work aims to provide a principled foundation for the next generation of LLM-enhanced recommender systems.

\section{A Perspective: From Alignment to Complementarity Fusion}
\label{sec:perspective}
In this section, we provide a formal abstraction of current collaborative–semantic alignment practices, analyze their underlying structural hypothesis, and propose a more general latent formulation that supports complementarity-aware fusion.
\subsection{The prevailing alignment paradigm}

Modern recommender systems increasingly rely on two types of item representations: collaborative embeddings learned from interaction data and semantic embeddings derived from content encoders such as BERT, CLIP, or LLMs. 
The common motivation is clear. Collaborative signals capture collective behavioral structure, while semantic encoders provide fine-grained content information. Combining both appears natural.

Most recent methods integrate these two views through representation alignment. 
Formally, for each item $i$, a collaborative model produces:
\begin{equation}
e^{\text{CF}}_i = f_{\text{CF}}(\mathcal{D}_{\text{beh}}),
\end{equation}
and a semantic encoder produces
\begin{equation}
e^{\text{SEM}}_i = f_{\text{SEM}}(x_i),
\end{equation}
where $\mathcal{D}_{\text{beh}}$ denotes behavioral data and $x_i$ denotes item content.

Lightweight projection heads are then introduced,
\begin{equation}
z^{\text{CF}}_i = g_{\text{CF}}(e^{\text{CF}}_i), \quad
z^{\text{SEM}}_i = g_{\text{SEM}}(e^{\text{SEM}}_i),
\end{equation}
where \(g_{\text{CF}}\) and \(g_{\text{SEM}}\) denote lightweight projection heads (e.g., linear layers or shallow MLPs) that map the collaborative and semantic embeddings into a common alignment space. These heads are trained with an alignment objective such as:
\begin{equation}
\mathcal{L}_{\text{align}}
= - \sum_i \log
\frac{\exp(\mathrm{sim}(z^{\text{CF}}_i, z^{\text{SEM}}_i)/\tau)}
{\sum_j \exp(\mathrm{sim}(z^{\text{CF}}_i, z^{\text{SEM}}_j)/\tau)},
\end{equation}
where \(\mathcal{L}_{\text{align}}\) is the item-level contrastive alignment objective over collaborative–semantic embedding pairs, \(\mathrm{sim}(\cdot,\cdot)\) denotes cosine similarity unless otherwise specified, and \(\tau > 0\) is a temperature parameter that controls the sharpness of the contrastive distribution.

Although implementations vary, the structural hypothesis is consistent: the two embedding spaces can be made geometrically compatible through a low-capacity transformation, typically linear or shallow. The training objective encourages the two spaces to become globally consistent.

\subsection{The alignment hypothesis}

We refer to this underlying belief as \textbf{the global low-complexity alignment hypothesis}. 
It assumes that each item has an underlying latent factor $z_i \in \mathbb{R}^k$, where $k$ denotes the latent dimensionality, and that collaborative and semantic embeddings are two parameterizations of the same latent structure:
\begin{equation}
e^{\text{CF}}_i = h_{\text{CF}}(z_i), \quad
e^{\text{SEM}}_i = h_{\text{SEM}}(z_i),
\end{equation}
where $h_{\text{CF}}$ and $h_{\text{SEM}}$ are view-specific parameterization functions that map the latent factor $z_i$ to the collaborative and semantic embedding spaces, respectively.

Under smoothness and low-complexity hypothesis, there exists a transformation $T: \mathbb{R}^{d_{\text{SEM}}} \rightarrow \mathbb{R}^{d_{\text{CF}}}$, typically a linear map or shallow MLP, such that
\begin{equation}
e^{\text{CF}}_i \approx T(e^{\text{SEM}}_i),
\end{equation}
or conversely in the other direction.

This formulation makes alignment conceptually simple. Multi-view integration becomes a problem of learning a shallow mapping between two spaces. If the hypothesis holds, stronger alignment implies better integration.

\subsection{A shared and private latent structure}

\begin{figure}[t!]
    \centering
    \includegraphics[width=\linewidth]{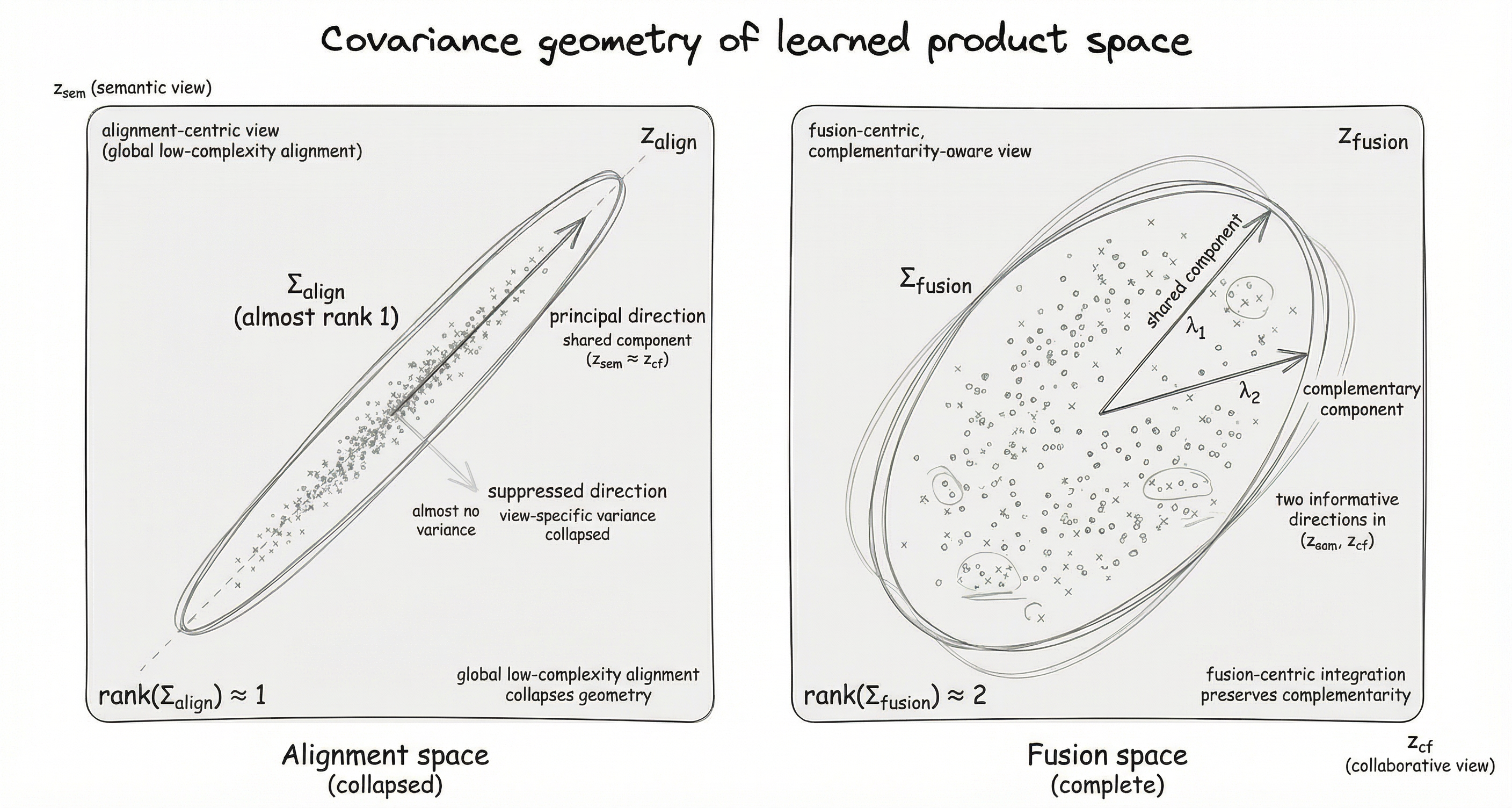}
\caption{\textbf{Schematic illustration of covariance geometry in the learned product space.}
Conceptual comparison between the alignment-centric view (left) and the proposed complementarity fusion-centric view (right).
While global alignment forces the representation geometry to collapse into a shared rank-1 subspace (suppressing view-specific variance), the complementarity fusion-centric approach preserves both shared ($\lambda_1$) and complementary ($\lambda_2$) components, maintaining the informational integrity of both collaborative and semantic views. Ellipses and axes are stylized for clarity and do not represent actual learned covariance matrices.}
    \label{fig:framework}
\end{figure}
We argue that this hypothesis is generally too strong for modern recommender systems. 
Collaborative and semantic embeddings are not merely alternative encodings of the same latent space.

A more realistic abstraction is a shared plus private latent structure. 
For each item $i$, let
\begin{equation}
z_i = \big(z^{\text{shared}}_i,\; z^{\text{CF-only}}_i,\; z^{\text{SEM-only}}_i\big).
\end{equation}

Collaborative embeddings are generated from
\begin{equation}
e^{\text{CF}}_i = g_{\text{CF}}\!\left(z^{\text{shared}}_i, z^{\text{CF-only}}_i\right),
\end{equation}
while semantic embeddings are generated from
\begin{equation}
e^{\text{SEM}}_i = g_{\text{SEM}}\!\left(z^{\text{shared}}_i, z^{\text{SEM-only}}_i\right).
\end{equation}

The shared component $z^{\text{shared}}_i$ captures stable semantic properties and long-term preference factors that are consistently manifested across both interaction data and content features. 
It encodes invariant structural information that reflects the intrinsic alignment between user interests and item semantics, independent of modality-specific noise or sampling artifacts. 
By modeling this shared latent space explicitly, the framework encourages cross-view consistency and facilitates knowledge transfer between collaborative and semantic representations.

The collaborative-specific component $z^{\text{CF-only}}_i$ models signals that arise purely from user--item interaction behavior. 
These signals may include popularity dynamics, exposure bias, feedback loops, position effects, temporal trends, and collective behavioral regularities shaped by platform mechanisms. 
Such patterns often reflect short-term or system-induced effects rather than intrinsic item characteristics, and therefore may not be directly recoverable from textual descriptions alone. 
Isolating this component allows the model to capture interaction-driven phenomena without contaminating the shared semantic structure.

In contrast, the semantic-specific component $z^{\text{SEM-only}}_i$ captures information derived from textual or other content-based modalities. 
This component may encode stylistic variation, nuanced content attributes, contextual semantics, descriptive richness, and fine-grained properties that are only weakly observed in sparse implicit feedback data. 
Semantic signals can provide complementary structural cues, especially in cold-start or long-tail scenarios where interaction evidence is limited. 

By decomposing representations into shared and modality-specific components, the model achieves a structured disentanglement between invariant preference factors and complementary modality-dependent information. 
This decomposition not only improves interpretability, but also enhances robustness by reducing interference across heterogeneous data sources.
Under this generative structure, recovering one full space from the other is not well-defined. Each view contains information that is not observable in the other. Global alignment therefore requires compressing private factors into an artificial shared geometry, which may distort both spaces, as illustrated in the left panel of Figure~\ref{fig:framework}.

\subsection{Why global alignment may be mismatched?}

If the two spaces are only partially shared, full-space alignment is a questionable objective. Forcing geometric consistency can suppress view-specific factors that do not match across views, even when they are predictive for ranking. In sparse regimes, interaction embeddings may also be noisy, and aligning semantic embeddings to such signals risks distorting semantic structure and reducing robustness. Moreover, global alignment blends behavioral and content-driven factors without explicitly modeling their differences, which can make both representations harder to interpret and control. We do not argue against alignment in general; rather, we caution against treating global low-complexity alignment as the default integration principle when the two views contain substantial private factors. These considerations suggest that agreement should not be the default objective in multi-view recommendation.

We therefore advocate a shift in emphasis: instead of asking how to globally align collaborative and semantic spaces, the more fundamental question is how to combine complementary signals while preserving their structural differences. We outline three directions:
\begin{itemize}[leftmargin=*]
    \item \textbf{Factorized integration.} Model shared and view-specific components explicitly, and align only the shared factors.
    \item \textbf{Late fusion as a principled baseline.} When structural compatibility is uncertain, representation-level or score-level fusion combines evidence without enforcing geometric collapse.
    \item \textbf{Complementarity-aware evaluation.} Quantify redundancy and complementarity between views, rather than relying solely on overall ranking metrics.
\end{itemize}

\subsection{Core Proposed Claims}

We summarize our perspective through three claims:
\begin{itemize}[leftmargin=*]
    \item \textbf{Claim 1 (Complementarity).} Collaborative and semantic representations overlap only partially; each encodes substantial view-specific factors, leading to low item-level agreement and strong ``unique-hit'' contributions across views.
    \item \textbf{Claim 2 (Limits of low-complexity alignment).} The premise that a low-capacity mapping can \emph{globally} align semantic and collaborative spaces is often unreliable: it may fit shared structure but fail to recover the full collaborative geometry, especially under distribution shifts.
    \item \textbf{Claim 3 (Paradigm shift).} Recommender systems should move beyond alignment-centric integration and explicitly model the complementarity between collaborative and semantic views, which our experiments show contributes a larger yet previously overlooked performance gain than alignment alone.
\end{itemize}

\section{Empirical Investigation}
\label{sec:empirical_study}

To validate our perspective on the topological dissonance between collaborative and semantic spaces, we conduct a series of systematic experiments across multiple recommendation settings.

\subsection{Experimental Setup}
\label{sec:exp_setup}

\subsubsection{Datasets}
We utilize three sparse recommendation benchmarks derived from \textbf{Amazon Reviews'23} \cite{hou2024bridging}: \textsc{Movies}, \textsc{Books}, and \textsc{Games}.
We apply standard 5-core filtering. As shown in Table~\ref{tab:datasets}, all datasets exhibit extreme sparsity ($>99.9\%$), making them ideal testbeds for evaluating long-tail signal preservation \textbf{under severe interaction scarcity}.
For the semantic view, we construct item documents by concatenating metadata fields (Title, Category, Description) and encode them using \textbf{BGE-M3} \cite{bge-m3}.

\begin{table}[t!]
    \centering
    \caption{Statistics of the datasets. The extreme sparsity highlights the challenge of pure collaborative learning.}
    \label{tab:datasets}
    \resizebox{\linewidth}{!}{%
    \begin{tabular}{lcccc}
    \toprule
    \textbf{Dataset} & \textbf{\#Users} & \textbf{\#Items} & \textbf{\#Interactions} & \textbf{Sparsity} \\
    \midrule
    \textsc{Movies} & 27,292 & 24,608 & 331,049 & 99.95\% \\
    \textsc{Books} & 70,679 & 23,777 & 815,780 & 99.95\% \\
    \textsc{Games} & 43,430 & 11,767 & 383,477 & 99.92\% \\
    \bottomrule
    \end{tabular}%
    }
\end{table}

\subsubsection{Independent View Encoders}
\label{sec:backbones}

To strictly investigate the intrinsic properties of collaborative and semantic spaces without mutual interference, we construct and train two \textbf{independent} models as proxies for the two views.

\begin{itemize}[leftmargin=*]
    \item \textbf{Collaborative Baseline (CF-Only):} We employ standard \textbf{LightGCN} \cite{lightgcn} as the representative for the collaborative view. It is trained solely on interaction data using the BPR loss to capture pure structural signals. The embedding dimension is $d=64$.
    
    \item \textbf{Semantic Baseline (Sem-Only):} We construct a pure content-based model using \textbf{BGE-M3} \cite{bge-m3}. Item texts are encoded into 1024-d vectors and projected to $d=64$ via a linear layer. User representations are derived by mean-pooling the projected embeddings of their interaction history. Crucially, this model is trained with an InfoNCE objective ($\tau=0.15$) using global negative sampling, relying \textit{exclusively} on semantic features without accessing the collaborative graph structure during optimization.
    
    \item \textbf{Training Protocols:} Both models are implemented in PyTorch and optimized \textbf{independently}. We use the \textbf{Adam} optimizer (LR=$1e^{-3}$, Batch Size=2048, Weight Decay=$1e^{-4}$). We evaluate the model every 5 epochs, and training is terminated via early stopping if Recall@20 on the validation set does not improve for \textbf{5 consecutive evaluations}.
\end{itemize}
\subsubsection{Diagnostic Protocols}
\label{sec:metrics}
Beyond standard ranking metrics (Recall@K, NDCG@K), we formalize three specific metrics to diagnose the \textit{alignment-complementarity trade-off}. Let $\mathcal{U}$ denote the set of evaluation users, and let $\mathcal{H}^{(A)}_u$ and $\mathcal{H}^{(B)}_u$ denote the sets of correct hits (items in test set) in the Top-$K$ recommendations from model $A$ and model $B$, respectively.

\paragraph{1. Overlap Analysis (HitJaccard).}
To measure how much the two views mimic each other's successful predictions, we compute the Jaccard similarity of their hit sets:
\begin{equation}
    \mathrm{Jaccard}@K = \frac{1}{|\mathcal{U}|} \sum_{u \in \mathcal{U}} \frac{|\mathcal{H}^{(A)}_u \cap \mathcal{H}^{(B)}_u|}{|\mathcal{H}^{(A)}_u \cup \mathcal{H}^{(B)}_u|}.
\end{equation}
A lower Jaccard score indicates that the views are solving different subsets of user needs.

\paragraph{2. Complementarity Ratio (CompRatio).}
We propose \textit{CompRatio} to explicitly quantify the proportion of "unique gains" obtained by fusing two views. It is defined as the ratio of the symmetric difference (unique hits) to the total hits:
\begin{equation}
    \mathrm{CompRatio}@K = \frac{1}{|\mathcal{U}|} \sum_{u \in \mathcal{U}} \frac{|\mathcal{H}^{(A)}_u \triangle \mathcal{H}^{(B)}_u|}{|\mathcal{H}^{(A)}_u \cup \mathcal{H}^{(B)}_u|},
\end{equation}
where $\triangle$ denotes the symmetric difference ($\mathcal{H}^{(A)}_u \cup \mathcal{H}^{(B)}_u \setminus \mathcal{H}^{(A)}_u \cap \mathcal{H}^{(B)}_u$). High CompRatio implies strong complementarity.

\paragraph{3. Union-TopK Reference Upper Bound (UUB)}
To estimate the theoretical performance limit of fusion without improving individual encoders, we construct a Union-TopK Reference Upper Bound (UUB), which measures the maximum exploitable hit potential contained in the union of two single-view candidate lists:
\begin{equation}
\mathrm{UUB@}K = \frac{1}{|\mathcal{U}|} \sum_{u \in \mathcal{U}} \frac{\left| \mathcal{T}_u \cap \left(\mathcal{L}_u^{(A)} \cup \mathcal{L}_u^{(B)}\right) \right|}{|\mathcal{T}_u|},
\end{equation}
where $\mathcal{T}_u$ denotes the ground-truth relevant item set (i.e., test set items) for user $u$, and $\mathcal{L}_u^{(A)}$, $\mathcal{L}_u^{(B)}$ are the Top-$K$ recommendation lists produced by view $A$ and view $B$, respectively. Comparing the fused model against this bound reveals how much complementary potential is lost during alignment-based fusion and helps diagnose whether fusion genuinely introduces new information rather than merely re-ranking existing candidates.

\subsection{E1: Complementarity Diagnostics Between Semantic and Collaborative Views}
\label{sec:exp_complementarity}

This experiment aims to validate \textbf{Claim 1}: that collaborative and semantic representations overlap only partially and encode substantial view-specific factors. We establish this by diagnosing the behavior of two independent, clean baselines, one purely collaborative and one purely semantic, in a controlled setting without any forced alignment or complex fusion.

\subsubsection{Single-view Baseline Performance}
\label{sec:exp_complementarity_perf}

We first establish that both single-view models are reasonable and comparable baselines, ensuring that any observed differences are due to structural properties rather than a performance gap (i.e., avoiding a "strawman" comparison).
The collaborative baseline is a standard LightGCN ($d=64$). The semantic baseline is a simple linear probe ($d=64$) on frozen BGE-M3 embeddings, optimized via InfoNCE to retrieve items based on the mean-pooled semantic user history.

As shown in Table~\ref{tab:single_view_perf}, while LightGCN generally outperforms the semantic baseline, the performance difference is relatively narrow. For instance, on the Movies dataset, the semantic model actually achieves a slightly higher Recall@20 (0.1136 vs. 0.1100). This confirms that both views possess strong, independent predictive power, setting a fair stage for complementarity analysis.

\begin{table}[t!]
    \centering
    \caption{Performance comparison of single-view baselines. Both models achieve comparable performance levels, ensuring a fair diagnostic comparison.}
    \label{tab:single_view_perf}
    \begin{tabular}{l|cc|cc}
        \toprule
        \multirow{2}{*}{\textbf{Dataset}} & \multicolumn{2}{c|}{\textbf{LightGCN (CF)}} & \multicolumn{2}{c}{\textbf{Semantic (Content)}} \\
         & Recall@20 & Hit@20 & Recall@20 & Hit@20 \\
        \midrule
        Movies & 0.1100 & 0.1254 & 0.1136 & 0.1283 \\
        Books & 0.1902 & 0.2137 & 0.1743 & 0.1948 \\
        Games & 0.1755 & 0.1849 & 0.1659 & 0.1754 \\
        \bottomrule
    \end{tabular}
\end{table}

\subsubsection{Item-level Agreement: Overlap, Unique Hits, and Complementarity Ratio}
\label{sec:exp_complementarity_overlap}

We quantify the structural divergence between the two views by analyzing their Top-20 recommendation lists. We employ three metrics: List Jaccard (similarity of recommended lists), Hit Jaccard (similarity of correct hits), and Complementarity Ratio (the proportion of unique hits from either model relative to total hits).

The results in Table~\ref{tab:complementarity_overlap} reveal a striking phenomenon: despite having similar overall performance, the two models solve significantly different subsets of user needs.
\begin{itemize}[leftmargin=*]
    \item \textbf{Low Overlap:} The List Jaccard similarity is remarkably low across all datasets (e.g., 6.37\% on Movies), indicating that the recommendation landscapes of the two models are largely disjoint.
    \item \textbf{High Complementarity:} The Complementarity Ratio consistently exceeds 55\%. On the Movies dataset, \textbf{63.03\%} of all correct hits are unique to either the CF or Semantic model. This high proportion of view-specific hits implies that the semantic view captures distinct user preferences that would likely be missed if forced to align with the collaborative view.
\end{itemize}

These findings suggest that forcing geometric alignment (i.e., minimizing the distance between $z_{CF}$ and $z_{SEM}$) risks "homogenization," effectively discarding these view-specific hits (over 60\% of successful recommendations) that the semantic model uniquely captures, ultimately reducing diversity in recommendations.
\begin{table}[t!]
    \centering
    \small
    \renewcommand{\arraystretch}{0.9}
    \setlength{\tabcolsep}{5pt}
    
    \caption{Complementarity diagnostics. Extremely low overlap and high complementarity ratio (>55\%) indicate that the two views capture largely orthogonal signals.}
    \label{tab:complementarity_overlap}
    
    \begin{tabular}{l|cc|c}
        \toprule
        \textbf{Dataset} & \textbf{List Overlap} & \textbf{Hit Overlap} & \textbf{Comp. Ratio} \\
         & \footnotesize \textit{Jaccard@20} & \footnotesize \textit{Jaccard@20} & \footnotesize \textit{(Item-Level)} \\
        \midrule
        Movies & 6.37\% & 6.82\% & \textbf{63.03\%} \\
        Books & 14.53\% & 12.34\% & 56.43\% \\
        Games & 19.20\% & 10.78\% & 57.05\% \\
        \bottomrule
    \end{tabular}
\end{table}

\subsubsection{Confidence and Popularity Effects}
\label{sec:exp_complementarity_popularity}
We further dissect complementarity by examining how it varies with recommendation confidence (Top-$K$) and item popularity.
 
\textbf{High Confidence Complementarity.} Figure 2(d) shows that the Complementarity Ratio peaks at $K=5$ (66\%--69\%), indicating maximum divergence in high-confidence predictions. As $K$ increases, Figure 2(a) shows that list overlap rises (e.g., Jaccard increases), suggesting models converge only when retrieving generic items outside their core competencies. This proves fusion is most critical for the top ranks.
 
\textbf{Semantic View Favors the Long Tail.} We analyze the average Log-Popularity of unique hits. As shown in Figure 2(b), CF-unique hits (Blue) consistently target the head. In contrast, Semantic-unique hits (Red) exhibit a sharp popularity drop (e.g., -18.4\% on Movies, -11.1\% on Books). Note that even the smaller log-scale drop on Games (-6.3\%) implies a substantial reduction in raw interaction counts, confirming that the semantic view is a crucial long-tail compensator that alignment would suppress.




\begin{figure}[t!]
    \centering
    \includegraphics[width=\linewidth]{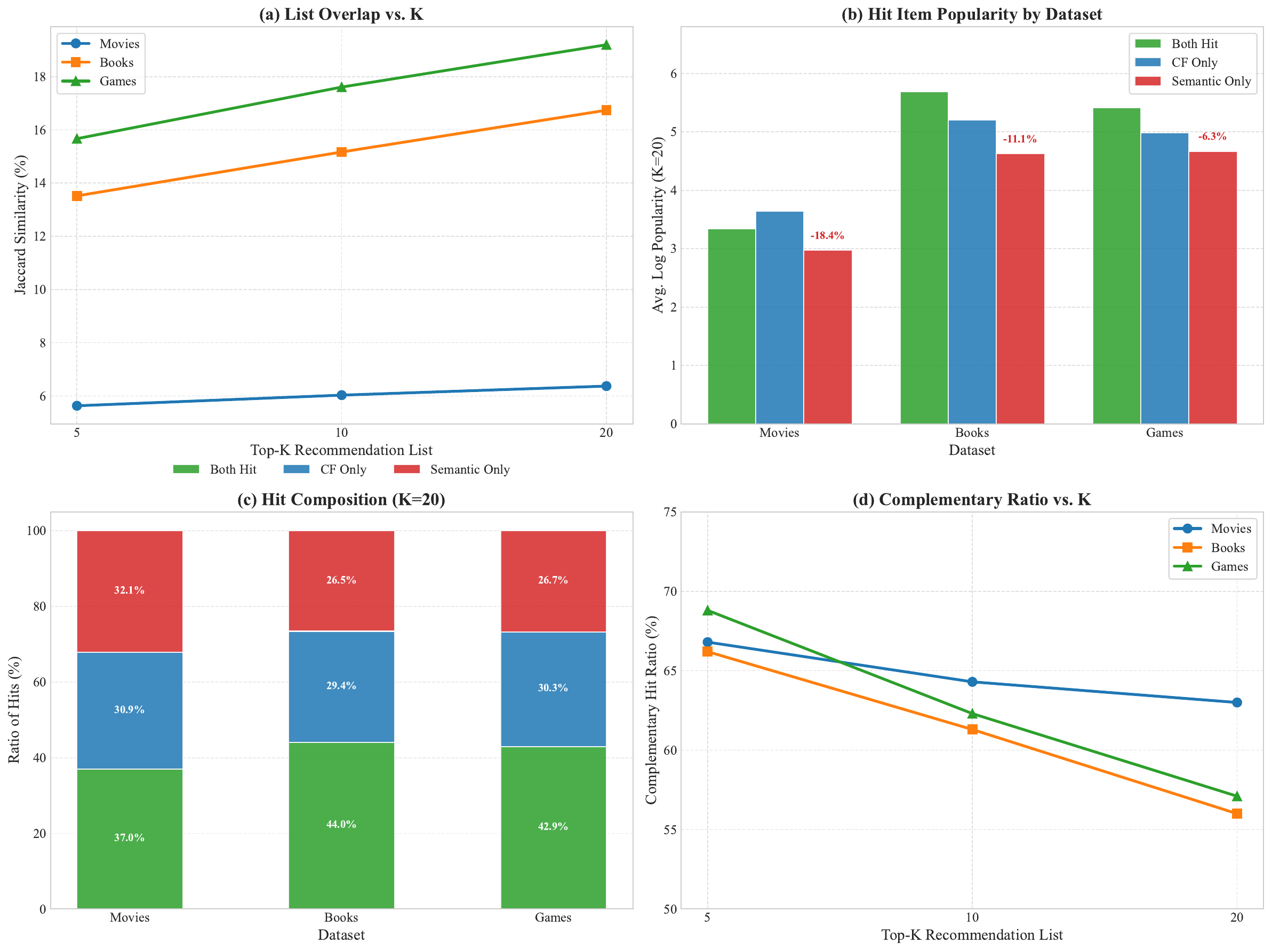}
    \caption{Complementarity Diagnostics. (a) Low List Overlap across all K indicates distinct ranking logic. (b) \textbf{The ``Pop. Drop''}: Unique hits from the Semantic model (Red) have significantly lower popularity than CF hits (Blue), confirming its advantage in the long tail. (c) Unique hits account for >60\% of total hits. (d) Complementarity Ratio peaks at low K, showing that the two views are most divergent in their high-confidence predictions.}
  \label{fig:topk_comp}
\end{figure}
\subsection{E2: One-way Alignment Probe Under Favorable Conditions}
\label{sec:exp_alignment_probe}

To test \textbf{Claim 2}, we probe whether a low-capacity global mapping can recover collaborative geometry from semantic embeddings under deliberately favorable conditions. We freeze both encoders and learn only a projection head from semantic to collaborative space; if alignment is still weak or fails to generalize, this supports a genuine structural mismatch rather than an optimization artifact.

\subsubsection{Protocol: Fix Encoders, Learn Only a Mapping}
\label{sec:probe_protocol}

We use converged LightGCN item embeddings ($d=64$) as targets $e^{CF}_i$,
and BGE-M3 item embeddings projected to $d=64$ as inputs $e^{SEM}_i$.
We train a mapping $T(\cdot)$ on a random 80\% subset of items
$\mathcal{I}_{\text{train}}$ by minimizing:
\begin{equation}
    \mathcal{L}_{probe}=\sum_{i\in \mathcal{I}_{\text{train}}}
    \|T(e^{SEM}_i)-e^{CF}_i\|_2^2 .
\end{equation}
We report $R^2$, Cos, GeoJac, RankCor, ListJac, and downstream Recall.

\subsubsection{Results: Capacity and Generalization}
\label{sec:probe_results}

We sweep $T$ from Linear to shallow/deeper MLPs and evaluate the alignment
probe using seven metrics: $R^2$ (coefficient of determination, measuring
how well the mapped semantic embeddings explain the variance of the target
collaborative embeddings), \textbf{Cos} (average cosine similarity between
mapped and target embeddings), \textbf{GeoJac} (geometric-neighborhood
Jaccard similarity~\cite{niwattanakul2013using}, measuring the overlap of nearest-neighbor
sets in the original and mapped spaces), \textbf{RankCor} (Spearman rank
correlation between neighbor rankings induced by the two spaces),
\textbf{ListJac} (Jaccard similarity between Top-$K$ neighbor lists
retrieved from each space), \textbf{Recall(CF)} (Recall@20 achieved by the
target collaborative embeddings, i.e., the performance ceiling), and
\textbf{Recall(Ps)} (Recall@20 achieved by the projected semantic embeddings
after applying the learned mapping $T(\cdot)$). On \textbf{training items}
(Table~\ref{tab:probe_train}), low-capacity mappings fit poorly (e.g.,
Linear $R^2{=}0.171$, ListJac $0.039$), while a large MLP can nearly match the target, indicating strong capacity dependence.

However, on \textbf{unseen
test items} (Table~\ref{tab:probe_test}), performance degrades sharply; even high-capacity models exhibit negative $R^2$ and substantially lower Recall than the collaborative upper bound, suggesting that they largely memorize training correspondences rather than learning a genuinely transferable geometric relationship between the two representation spaces.

\subsubsection{Takeaway}
\label{sec:probe_takeaway}

Even under frozen-encoder, low-interference settings, collaborative structure is not reliably recoverable from semantics via a low-capacity global mapping. This supports the view that the collaborative space contains substantial private factors that are not encoded in the semantic view, making global alignment an unsafe default assumption.

\begin{table}[t!]
    \centering
    \caption{Alignment probe results on \textbf{Training Items}. Low-capacity
    mappings (Linear, MLP-0) fit poorly; high-capacity models can force
    alignment on seen items.}
    \label{tab:probe_train}
    \resizebox{\linewidth}{!}{
    \begin{tabular}{l|ccccc|cc}
        \toprule
        \textbf{Model} & $\mathbf{R^2}$ & \textbf{Cos} & \textbf{GeoJac}
        & \textbf{RankCor} & \textbf{ListJac} & \textbf{Recall(CF)}
        & \textbf{Recall(Ps)} \\
        \midrule
        Identity & -0.076 & 0.015 & 0.042 & 0.162 & 0.001 & 0.1095
        & 0.0012 \\
        Linear Map & 0.171 & 0.420 & 0.031 & 0.312 & 0.039 & 0.1095
        & 0.0288 \\
        MLP-0 (Small) & 0.308 & 0.552 & 0.051 & 0.442 & 0.102 & 0.1095
        & 0.0655 \\
        MLP-1 (1-Hidden) & 0.475 & 0.686 & 0.086 & 0.603 & 0.178
        & 0.1095 & 0.0905 \\
        MLP-2 (2-Hidden) & \textbf{0.977} & \textbf{0.988}
        & \textbf{0.706} & \textbf{0.985} & \textbf{0.767} & 0.1095
        & \textbf{0.1095} \\
        \bottomrule
    \end{tabular}
    }
\end{table}

\begin{table}[t!]
    \centering
    \caption{Alignment probe results on \textbf{Test Items}. The failure to
    generalize (negative $R^2$, reduced Recall) indicates semantics are
    insufficient to reconstruct unseen collaborative signals.}
    \label{tab:probe_test}
    \resizebox{\linewidth}{!}{
    \begin{tabular}{l|ccccc|cc}
        \toprule
        \textbf{Model} & $\mathbf{R^2}$ & \textbf{Cos} & \textbf{GeoJac}
        & \textbf{RankCor} & \textbf{ListJac} & \textbf{Recall(CF)}
        & \textbf{Recall(Ps)} \\
        \midrule
        Linear Map & 0.166 & 0.415 & 0.044 & 0.322 & 0.039 & 0.1005
        & 0.0251 \\
        MLP-0 (Small) & 0.161 & 0.423 & 0.052 & 0.307 & 0.097 & 0.1005
        & 0.0473 \\
        MLP-1 (1-Hidden) & 0.013 & 0.365 & 0.040 & 0.250 & 0.162
        & 0.1005 & 0.0549 \\
        MLP-2 (2-Hidden) & -0.487 & 0.290 & 0.025 & 0.166 & 0.504
        & 0.1005 & 0.0564 \\
        MLP-3 (3-Hidden) & -0.495 & 0.291 & 0.028 & 0.175 & 0.485
        & 0.1005 & 0.0595 \\
        \bottomrule
    \end{tabular}
    }
\end{table}

\subsection{E3: Minimal Fusion Turns Complementarity into Gains}

\subsubsection{Fusion Baseline: A "Norm-Concat-Norm" Design}
\label{sec:fusion_baseline}

To validate that simple fusion effectively exploits complementarity, we design a minimal dual-encoder framework enforcing strict constraints to prevent the collaborative view from dominating.

\paragraph{Decoupled Encoders \& Fusion Strategy.}

We train two independent branches: a \textbf{Semantic View} built on frozen
BGE-M3 features and projected to $d{=}64$ by a single linear layer to preserve
intrinsic topology, and a \textbf{Collaborative View} that starts from randomly
initialized user/item ID embeddings and applies LightGCN-style propagation on
the user--item bipartite graph ($L{=}2$) to capture pure interaction signals.

Let $\mathbf{x}_i \in \mathbb{R}^{d_s}$ denote the frozen BGE-M3 feature of
item $i$, and let
\begin{equation}
    \mathbf{e}_u^{SEM}=\mathrm{Norm}\!\left(\frac{1}{|N_u|}\sum_{i\in N_u}\mathbf{x}_i\right)
\end{equation}
be the user-level semantic input obtained by mean-pooling the interacted items.
Let $W_s \in \mathbb{R}^{d \times d_s}$ and $b_s \in \mathbb{R}^{d}$ be the
learnable projection parameters. On the collaborative side, we write
$\mathrm{GCN}(u)$ for the layer-averaged user embedding produced by the
collaborative graph encoder. We then adopt a ``Norm-Concat-Norm'' strategy,
applying $\ell_2$ normalization $\mathrm{Norm}(\cdot)$ both before and after
concatenation ($\oplus$):
\begin{align}
    \mathbf{h}_u^{sem} &= \mathrm{Norm}(W_s \mathbf{e}_u^{SEM} + b_s),\quad
    \mathbf{h}_u^{cf} = \mathrm{Norm}(\mathrm{GCN}(u)) \\
    \mathbf{z}_u &= \mathrm{Norm}(\mathbf{h}_u^{cf} \oplus \mathbf{h}_u^{sem})
\end{align}
Item-side representations are constructed analogously and used for scoring.
Pre-normalization ensures equal-magnitude contributions ($\|\cdot\|=1$) from
both views, preventing the easier-to-fit collaborative signal from dominating
the fusion pool.

\paragraph{Optimization with Dynamic Hard Negatives.}
We optimize the model end-to-end with an InfoNCE objective augmented by
\textbf{Dynamic Hard Negative Mining}. In each batch, negatives are selected as
the highest-scoring items under the current \textit{fused} representation. This
encourages cross-view cooperation: when collaborative signals are less
discriminative (e.g., under sparse interactions), the resulting gradients push
the semantic branch to resolve hard negatives, leading the two views to
complement each other.

\subsubsection{Overall Results: The Triumph of Fusion}
\label{sec:fusion_overall}

To rigorously test Claim 3, we first benchmark our minimal fusion strategy against two dominant paradigms: (1) \textbf{ID-based CF} (LightGCN, SimGCL, NCL) and (2) \textbf{Alignment/Mapping methods} (CARec, RLMRec, AlphaRec). 
Table~\ref{tab:main_performance} presents the comprehensive comparison. \textbf{The results reveal a startling trend:} our Naive Fusion strategy---despite lacking complex alignment losses or generative fine-tuning---consistently outperforms all baselines across all datasets. Notably, it surpasses \textbf{RLMRec} and \textbf{CARec}, which rely on heavy "alignment" machinery. This empirical evidence directly challenges the necessity of geometric alignment, suggesting that preserving view-specific complementarity (Union) is more effective than enforcing agreement (Intersection). 

To dissect the source of these gains, we evaluate the internal Semantic and Collaborative views extracted from the trained fusion model. As shown in Table~\ref{tab:fusion_overall}, simple fusion consistently outperforms the best constituent single view and nearly matches the Union-TopK Reference Upper Bound on Movies, confirming that it successfully exploits the latent complementarity between views.

\begin{table*}[t]
\centering
\caption{\textbf{Paradigm Comparison: Fusion vs. Alignment.} 
We compare different paradigms on three benchmarks. 
Results are reported with both $K=10$ and $K=20$.}
\label{tab:main_performance}

\resizebox{0.95\textwidth}{!}{
\begin{tabular}{p{1.2cm} p{1.9cm}|cccc|cccc|cccc}
\toprule
\multirow{2}{*}{\textbf{Paradigm}} & \multirow{2}{*}{\textbf{Model}} 
& \multicolumn{4}{c|}{\textbf{Books}} 
& \multicolumn{4}{c|}{\textbf{Movies}} 
& \multicolumn{4}{c}{\textbf{Games}} \\
\cmidrule(lr){3-6} \cmidrule(lr){7-10} \cmidrule(lr){11-14}
& 
& R@10 & R@20 & N@10 & N@20
& R@10 & R@20 & N@10 & N@20
& R@10 & R@20 & N@10 & N@20 \\
\midrule

\multirow{3}{*}{\textit{ID-CF}}
& LightGCN\cite{lightgcn} 
& 0.1361 & 0.1891 & 0.0798 & 0.0936
& 0.0814 & 0.1100 & 0.0508 & 0.0583
& 0.1206 & 0.1766 & 0.0657 & 0.0800 \\

& SimGCL\cite{SimGCL} 
& 0.1430 & 0.1947 & 0.0834 & 0.0969
& 0.0830 & 0.1105 & 0.0510 & 0.0582
& 0.1301 & 0.1890 & 0.0716 & 0.0870 \\

& NCL\cite{NCL}
& 0.1441 & 0.1990 & 0.0858 & 0.1001
& 0.0822 & 0.1092 & 0.0513 & 0.0583
& 0.1275 & 0.1871 & 0.0696 & 0.0848 \\
\midrule
\multirow{3}{*}{\shortstack{\textit{Alignment/}\\\textit{Mapping}}}
& CARec\cite{CARec}
& 0.1388 & 0.1902 & 0.0742 & 0.0948
& 0.0798 & 0.1178 & 0.0481 & 0.0565
& 0.1212 & 0.1810 & 0.0629 & 0.0803 \\

& RLMRec\cite{RLMRec}  
& 0.1423 & 0.1998 & 0.0804 & 0.1021
& 0.0823 & 0.1210 & 0.0535 & 0.0601
& 0.1253 & 0.1850 & 0.0689 & 0.0846 \\

& AlphaRec\cite{AlphaRec}  
& \underline{0.1668} & \underline{0.2247} & \underline{0.0990} & \underline{0.1147}
& \underline{0.1146} & \underline{0.1507} & \underline{0.0699} & \underline{0.0794}
& \underline{0.1350} & \underline{0.2000} & \underline{0.0734} & \underline{0.0900} \\
\midrule

\textit{Fusion} & \textbf{Naive Fusion} 
& \textbf{0.1786} & \textbf{0.2404} & \textbf{0.1088} & \textbf{0.1248}
& \textbf{0.1231} & \textbf{0.1579} & \textbf{0.0811} & \textbf{0.0902}
& \textbf{0.1516} & \textbf{0.2177} & \textbf{0.0856} & \textbf{0.1026} \\

\bottomrule
\end{tabular}
}
\end{table*}


\begin{table}[t!]
    \centering
    \caption{Component Analysis \& UUB. The fused model approaches the UUB, particularly on Movies, indicating effective exploitation of complementarity.}
    \label{tab:fusion_overall}
    \resizebox{\linewidth}{!}{
    \begin{tabular}{l|c|cc|cc|c}
        \toprule
        \textbf{Dataset} & \textbf{Metric} & \textbf{Sem. View} & \textbf{Collab. View} & \textbf{Fused} & \textbf{UUB} & \textbf{Gain vs Best} \\
        \midrule
        \multirow{2}{*}{Books} & Recall@20 & 0.1618 & 0.2120 & \textbf{0.2404} & 0.2670 & +13.4\% \\
         & NDCG@20 & 0.0813 & 0.1083 & \textbf{0.1249} & - & +15.3\% \\
        \midrule
        \multirow{2}{*}{Movies} & Recall@20 & 0.1115 & 0.1108 & \textbf{0.1580} & 0.1594 & +41.7\% \\
         & NDCG@20 & 0.0576 & 0.0652 & \textbf{0.0902} & - & +38.3\% \\
        \midrule
        \multirow{2}{*}{Games} & Recall@20 & 0.1358 & 0.1795 & \textbf{0.2178} & 0.2396 & +21.3\% \\
         & NDCG@20 & 0.0612 & 0.0845 & \textbf{0.1026} & - & +21.4\% \\
        \bottomrule
    \end{tabular}
    }
\end{table}

\subsubsection{Popularity-stratified Analysis}
\label{sec:fusion_stratified}
To localize the gains, we stratify Recall@20 into Head (Top 20\%), Mid (Next 20\%), and Cold (Bottom 60\%) groups. Table~\ref{tab:stratified_recall} reveals a distinct complementarity profile. In the massive Cold tail, Fusion \textit{preserves} semantic dominance, avoiding the collapse of pure CF. Crucially, the Mid group exhibits strong \textit{synergy}, where Fusion outperforms the best internal single view by up to 36.8\% (Books), confirming effective integration across the entire spectrum.

\begin{table}[t!]
    \centering
    \caption{Recall@20 Breakdown by Item Popularity. Semantics help Cold/Tail, collaboration helps Hot/Head, and fusion balances both.}
    \label{tab:stratified_recall}
    \resizebox{\linewidth}{!}{
    \begin{tabular}{l|c|cc|cc}
        \toprule
        \textbf{Dataset} & \textbf{Group} & \textbf{Sem. View} & \textbf{Collab. View} & \textbf{Fused} & \textbf{UUB} \\
        \midrule
        \multirow{3}{*}{Books} & Cold & \textbf{0.0690} & 0.0399 & 0.0664 & 0.0865 \\
         & Mid & 0.0928 & 0.0919 & \textbf{0.1270} & 0.1412 \\
         & Hot & 0.1917 & 0.2720 & \textbf{0.2983} & 0.3285 \\
        \midrule
        \multirow{3}{*}{Movies} & Cold & \textbf{0.0880} & 0.0469 & 0.0877 & 0.1048 \\
         & Mid & 0.1162 & 0.1001 & \textbf{0.1449} & 0.1510 \\
         & Hot & 0.1129 & 0.1437 & \textbf{0.1924} & 0.1834 \\
        \midrule
        \multirow{3}{*}{Games} & Cold & \textbf{0.0830} & 0.0311 & 0.0594 & 0.0998 \\
         & Mid & 0.1051 & 0.0819 & \textbf{0.1233} & 0.1471 \\
         & Hot & 0.1529 & 0.2322 & \textbf{0.2719} & 0.2890 \\
        \bottomrule
    \end{tabular}
    }
\end{table}

\subsubsection{Where Do the Gains Come From?}
\label{sec:fusion_gain_source}

We further inspect hit composition by computing the List Jaccard between the two internal branch rankings and labeling each hit as Semantic-Unique, Collab-Unique, or Common. Table~\ref{tab:hit_composition} shows very low Jaccard ($<0.1$) and a large fraction of unique hits from each view, indicating that the fused model retains strong complementarity rather than collapsing the two branches into one mode.

\begin{table}[t!]
    \centering
    \caption{Hit Composition Analysis. Low List Jaccard and balanced unique hits indicate preserved complementarity within the fused model.}
    \label{tab:hit_composition}
    \resizebox{\linewidth}{!}{
    \begin{tabular}{l|c|ccc}
        \toprule
        \textbf{Dataset} & \textbf{List Jaccard} & \textbf{Sem. Unique} & \textbf{Collab. Unique} & \textbf{Common Hits} \\
        \midrule
        Books & 0.0931 & 21.44\% & 41.78\% & 36.79\% \\
        Movies & 0.0390 & 31.55\% & 32.68\% & 35.77\% \\
        Games & 0.0861 & 25.49\% & 43.58\% & 30.93\% \\
        \bottomrule
    \end{tabular}
    }
\end{table}
\subsection{Visualization: Manifold Mismatch and Topological Conflict}
\label{sec:viz_mismatch}

\begin{figure}[t]
    \centering
    \begin{subfigure}[b]{0.48\linewidth}
        \centering
        \includegraphics[width=\linewidth]{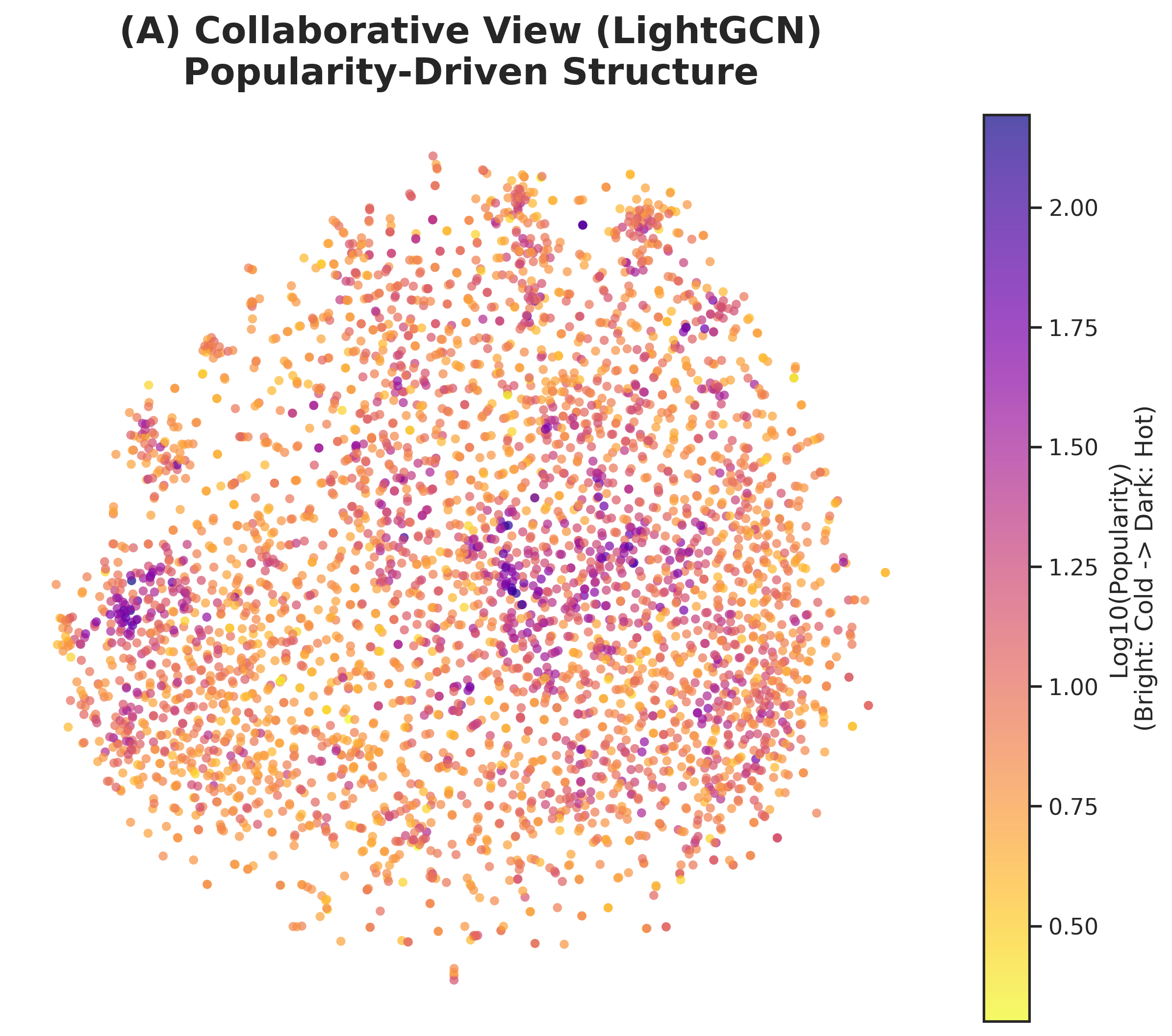}
        \caption{Collaborative View}
    \end{subfigure}
    \hfill
    \begin{subfigure}[b]{0.48\linewidth}
        \centering
        \includegraphics[width=\linewidth]{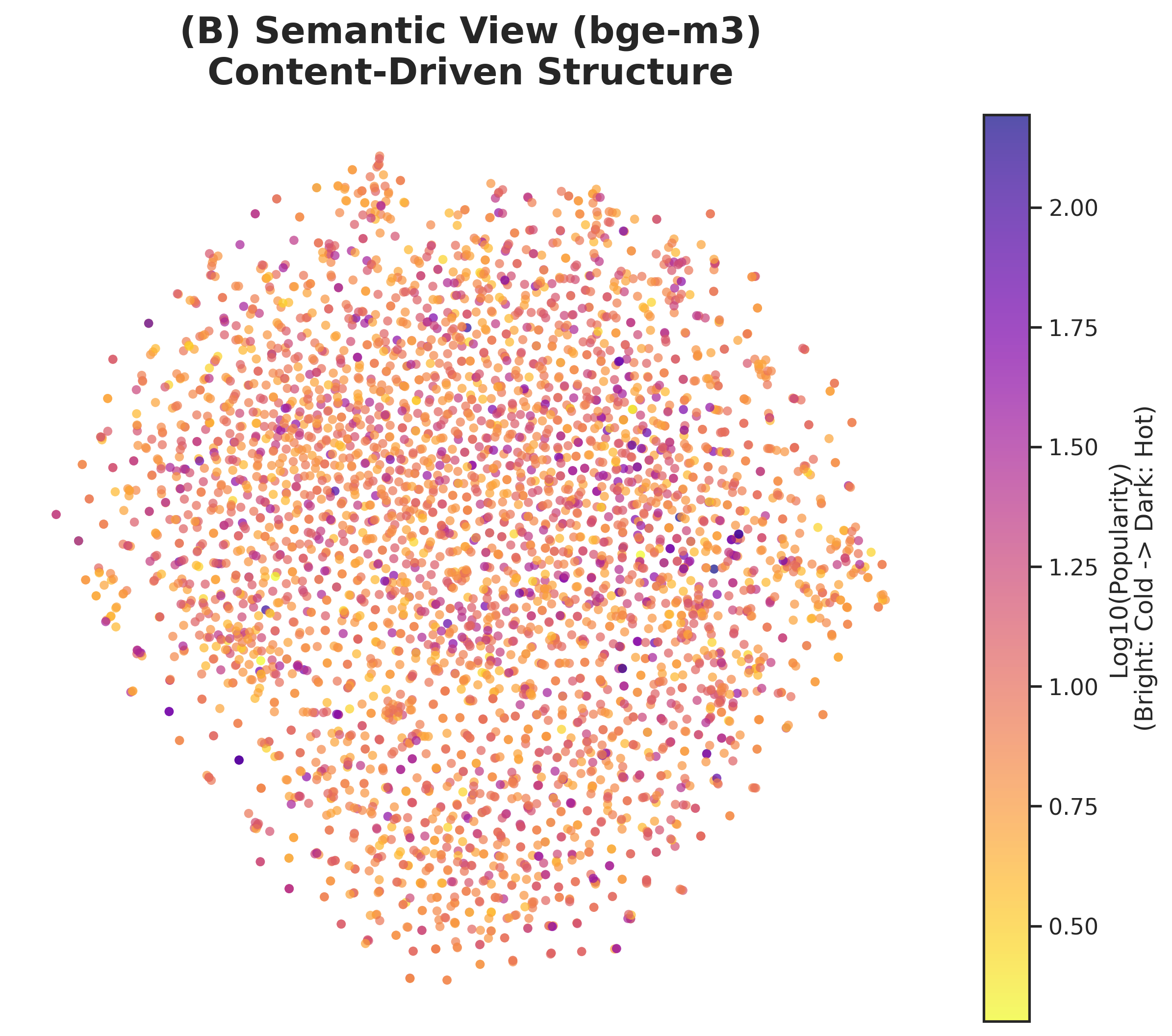}
        \caption{Semantic View}
    \end{subfigure}
    \caption{Global t-SNE visualization colored by Log-Popularity, from \textbf{Bright Yellow (Cold/Tail)} to \textbf{Dark Purple (Hot/Head)}. (a) The Collaborative View exhibits a ``popularity core'' where dark points cluster tightly in the center. (b) The Semantic View shows a more uniform distribution with mixed popularity gradients, indicating structure independent of interaction frequency.}
    \label{fig:global_tsne}
\end{figure}

\begin{figure}[t!]
    \centering
    \includegraphics[width=0.93\linewidth]{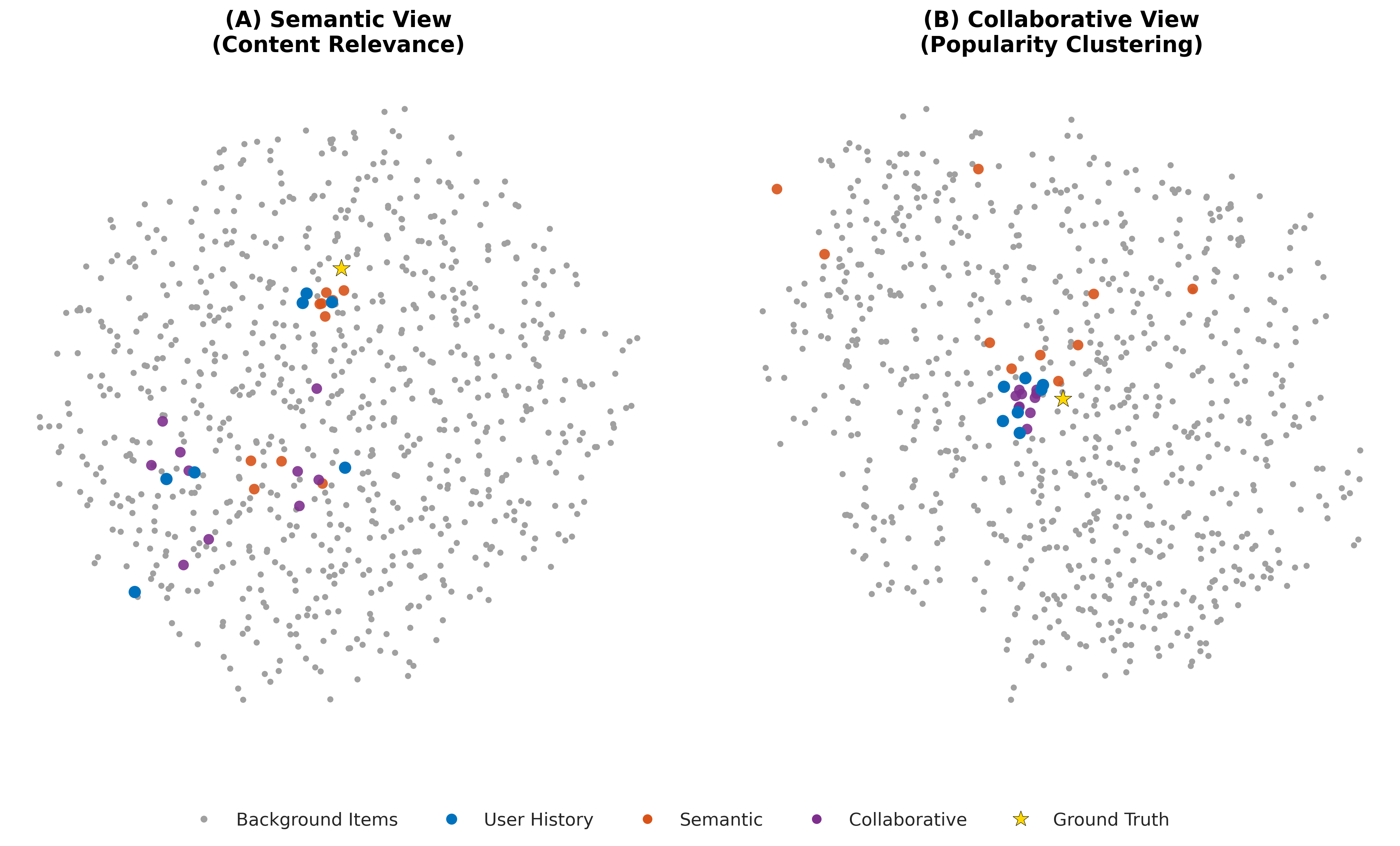}
    \caption{Local Universe Projection. (A) In the Semantic View, our recommendations (Red) align with user history (Blue), while Baseline (Purple) drifts. (B) In the Collaborative View, the Baseline collapses into the popularity center, while ours explores the sparse long-tail region.}
    \label{fig:local_universe}
\end{figure}

\begin{table}[t!]
    \centering
    \caption{Top-Neighbors Comparison: IP Consistency vs. Behavioral Generalization.}
    \label{tab:starwars_frozen}
    \resizebox{\linewidth}{!}{
    \begin{tabular}{c|l|l}
        \toprule
        \textbf{Rank} & \textbf{Semantic View (Proj)} & \textbf{Collaborative View (Pco)} \\
        \midrule
        \multicolumn{3}{c}{\textit{Anchor: Star Wars: The Force Awakens}} \\
        \midrule
        1 & Star Wars: The Force Awakens & Passengers [Blu-ray] \\
        2 & Star Wars: The Last Jedi & Ant-Man [Blu-ray] \\
        3 & STAR WARS: THE RISE OF SKYWALKER & Star Trek Beyond \\
        \midrule
        \multicolumn{3}{c}{\textit{Anchor: Frozen (Animation)}} \\
        \midrule
        1 & Frozen (Feature) [Blu-ray] & Frozen (Feature) \\
        2 & Frozen II & The Hunger Games: Mockingjay \\
        3 & Maleficent (Fantasy) & Despicable Me 2 \\
        \bottomrule
    \end{tabular}
    }
\end{table}

In this section, we provide qualitative evidence that the semantic space and the collaborative space are not structurally homomorphic, but instead exhibit clear geometric mismatch and complementary inductive biases. Rather than being simple reparameterizations of a shared latent structure, the two spaces organize items according to fundamentally different principles. All analyses are conducted on the \textbf{Movies} dataset. We use \textbf{LightGCN} to obtain collaborative embeddings learned purely from user--item interactions, and \textbf{BGE-M3} to obtain semantic embeddings derived from item metadata (title, category, description). This setup allows us to directly compare interaction-driven geometry and content-driven geometry under the same item universe.

\subsubsection{Macro-level: The Centripetal Force of Popularity}

We project both spaces to 2D using t-SNE to visualize their global organization. Figure~\ref{fig:global_tsne} reveals a continuous popularity-driven spectrum with sharply different structural arrangements in each view. In the \textbf{Collaborative Space}, popular items form a dense central core while tail items are scattered toward the periphery. This centripetal pattern reflects the influence of interaction frequency: high-degree items accumulate stronger co-occurrence signals and therefore occupy geometrically central positions. The resulting manifold is visibly shaped by popularity dynamics.

In contrast, the \textbf{Semantic Space} exhibits clusters organized primarily by content similarity rather than interaction intensity. Items group according to shared themes or descriptive attributes, and popularity colors are broadly mixed within clusters. The absence of a pronounced popularity core suggests that semantic embeddings preserve attribute-level structure with relative neutrality to exposure frequency. This contrast indicates that the two spaces encode different inductive biases—one driven by behavioral statistics, the other by intrinsic semantic coherence.
\begin{figure*}[t!]
  \centering
  \includegraphics[width=0.87\textwidth]{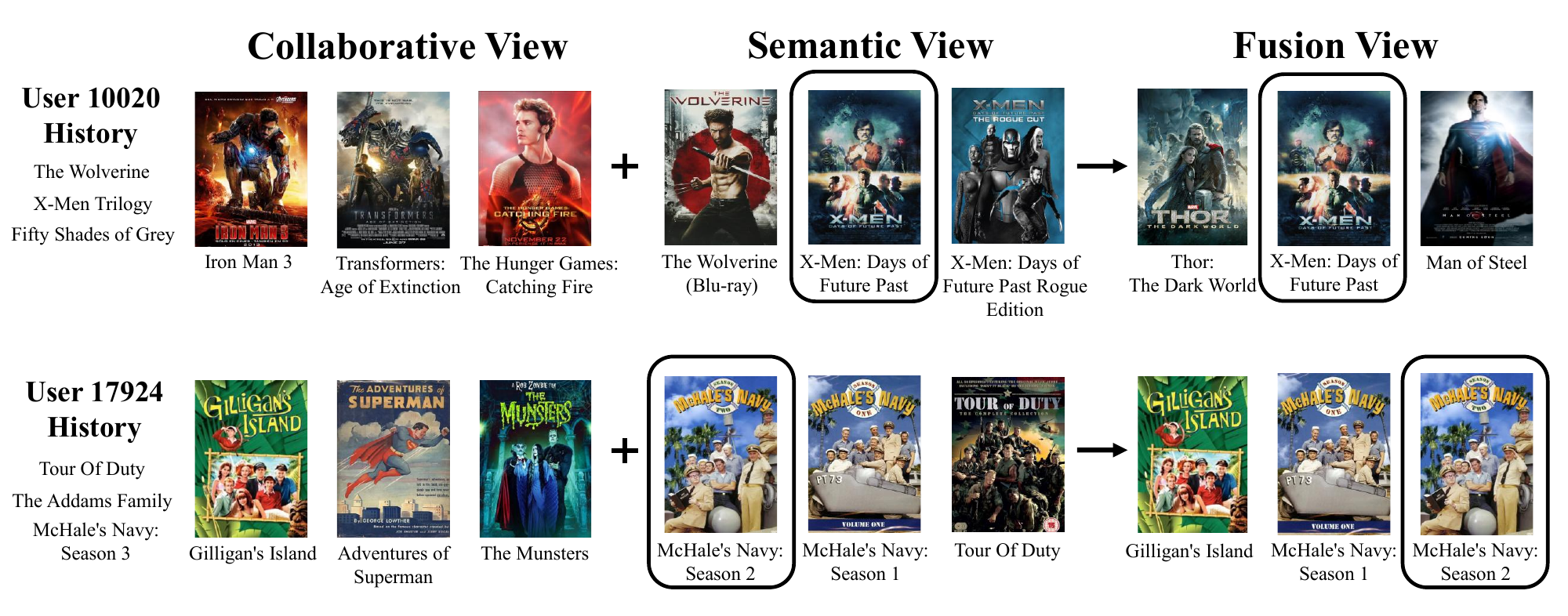}
\caption{%
    Top-3 recommendations from each view for two users.
    \textbf{User~10020} (X-Men fan): Collaborative retrieves generic blockbusters; Semantic stays within the franchise; Fusion combines both.
    \textbf{User~17924} (classic-TV viewer): Collaborative surfaces genre neighbours; Semantic identifies unseen seasons; Fusion retains both.
    Bold borders denote ground-truth test items.%
  }
  \label{fig:case_study_fusion}
\end{figure*}

\begin{figure}[t!]
    \centering
    \includegraphics[width=\linewidth]{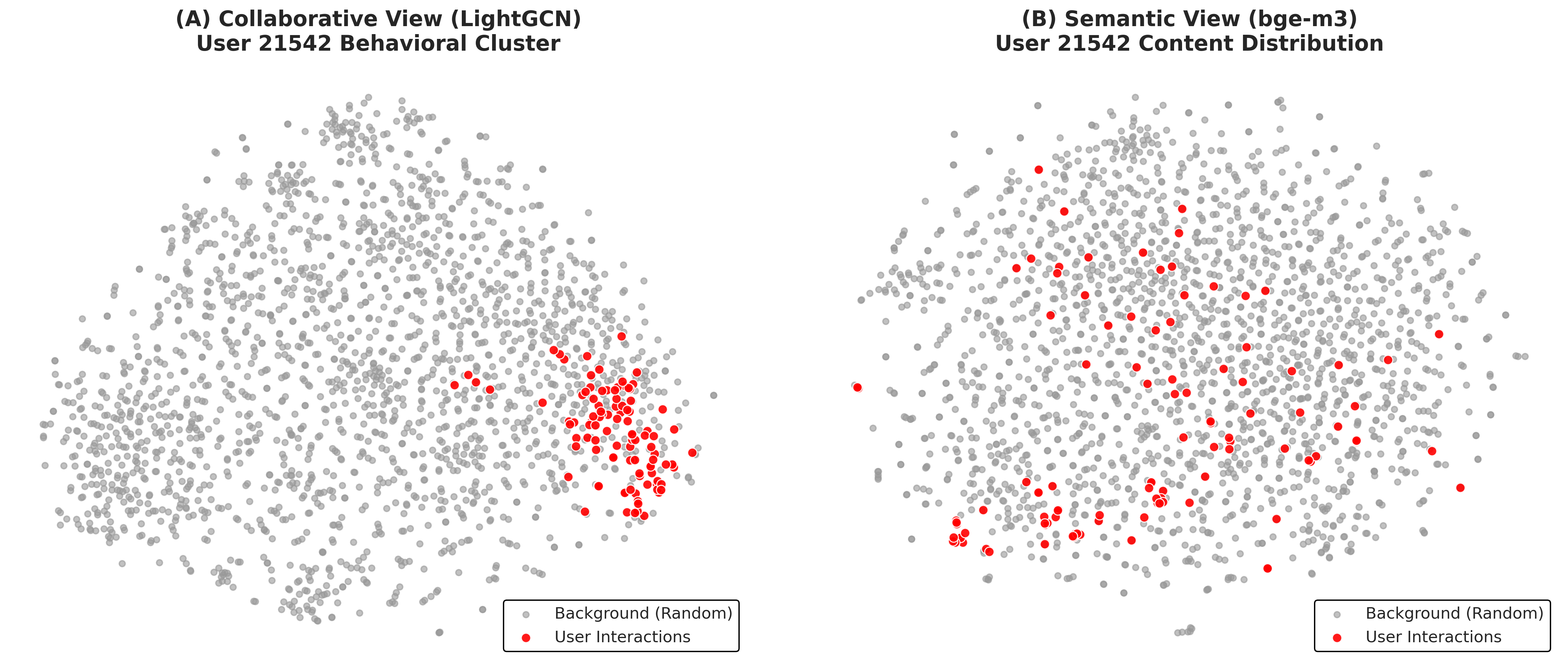}
    \caption{Visualization of High-Interaction Users. (A) The Collaborative View forms tight behavioral clusters, accurately capturing established user habits. (B) The Semantic View shows a more dispersed distribution, reflecting diverse content interests of active users.}
    \label{fig:warm_user}
\end{figure}

\subsubsection{Local Universe Projection: Visualizing Semantic Drift}

To connect global manifold mismatch to user-level behavior, we further visualize a warm-start user's local neighborhood. Figure~\ref{fig:local_universe} plots the user history (Blue), our semantic recommendations (Red), and the collaborative baseline recommendations (Purple) within each embedding space. This local perspective complements the global analyses by revealing how structural differences manifest in concrete retrieval outcomes.
In the semantic view, the user history forms a coherent thematic cluster, and our recommendations remain close to this region, indicating strong content-level continuity. By contrast, the collaborative baseline drifts away from the semantic cluster, reflecting interaction-driven signals not strictly aligned with textual similarity.

In the collaborative view, the baseline recommendations collapse toward the high-density popularity core, illustrating how popularity-shaped geometry can dominate ranking decisions and bias retrieval toward frequently interacted items. Meanwhile, our recommendations and the user history remain in sparser regions, suggesting that preserving semantic structure helps resist popularity-induced drift and maintains coverage over long-tail items. Together, these visualizations provide intuitive user-level evidence of topological conflict and motivate integration strategies that respect structural heterogeneity rather than enforcing uniform alignment.

\subsection{Case Study}
\label{sec:case_study}

We now zoom in from global visualization to concrete neighborhoods and user cases to examine how complementarity manifests in practice. The purpose is not to argue that one view is universally superior, but to demonstrate that they capture orthogonal yet meaningful facets of recommendation signals. Semantic embeddings emphasize explicit content similarity and entity-level coherence, whereas collaborative embeddings reflect implicit co-consumption patterns shaped by collective behavior.

\subsubsection{Micro-level: Content Anchoring vs. Behavioral Generalization}

Using \textit{Star Wars} and \textit{Frozen} as anchors, Table~\ref{tab:starwars_frozen} illustrates two distinct retrieval logics. In the semantic space, nearest neighbors remain tightly within the same intellectual property (IP) universe, retrieving sequels and closely related narrative extensions. This behavior reflects strong content anchoring and entity consistency.

In contrast, the collaborative space generalizes toward other blockbusters consumed by overlapping audiences, even when they belong to different franchises. This reflects demographic or audience-level similarity rather than narrative continuity. Neither pattern is inherently incorrect; rather, they correspond to different underlying questions. The semantic space encodes ``what the item is,'' while the collaborative space encodes ``who tends to consume the item.''

\subsubsection{Evidence for ``Union $>$ Intersection''}

The \textit{Real Steel} example (Table~\ref{tab:real_steel}) further clarifies why fusion can outperform alignment. The semantic view retrieves items sharing lexical cues such as ``Steel'' or ``Iron,'' regardless of genre distinctions. The collaborative view, however, surfaces latent thematic affinity—such as mecha or robot-related films—based on shared audience behavior.

If alignment forces geometric convergence, one of these signals may be suppressed. Fusion, by contrast, preserves the union of both perspectives: textual precision from semantics and latent structural insight from collaborative filtering. This illustrates why intersection-based integration can be unnecessarily restrictive.

\subsubsection{User Case Studies: Cold-Start Survival and Zero-Shot Retrieval}

To provide intuitive evidence for complementarity, Figure~\ref{fig:case_study_fusion} visualizes the Top-3 recommendations from each view for two representative users. The Collaborative View defaults to popularity-driven generalization (e.g., \emph{Iron Man~3}, \emph{Transformers} for User~10020), while the Semantic View anchors on content similarity (e.g., \emph{X-Men: Rogue Edition}, \emph{McHale's Navy: Season~1}). The Fusion View reconciles both signals: for User~10020 it retains the franchise-relevant \emph{X-Men: Days of Future Past} alongside thematically adjacent \emph{Thor: The Dark World}; for User~17924 it preserves season-level precision while incorporating \emph{Gilligan's Island} for broader coverage. This divergence is especially pronounced under cold-start conditions, where the collaborative baseline defaults to globally popular items due to sparse histories, while the semantic view maintains preference continuity by leveraging descriptive metadata and entity relationships. These examples illustrate that the fused list approximates the \emph{union} rather than the intersection of both views, consistent with our complementarity-preserving argument.

To provide a balanced perspective, we also visualize high inter-action users in Figure~\ref{fig:warm_user}. In this dense regime, the collaborative view forms tight behavioral clusters that closely reflect established consumption patterns, indicating that interaction signals are sufficiently rich to define stable preference geometry. By comparison, the semantic view appears more dispersed, as active users typically span multiple themes and content facets that are not easily captured by a single semantic cluster.

\subsubsection{Boundary of Complementarity: Collaborative Dominance}

Finally, we acknowledge a regime in which collaborative signals dominate. For users with rich interaction histories ($>50$ interactions), collaborative embeddings form tight behavioral clusters that accurately capture consumption patterns shaped by collective trends. In such dense regimes, semantic neighborhoods may appear more dispersed due to diverse interests and trend-driven behavior.
This boundary condition reinforces our core argument: semantic signals are particularly valuable for cold-start and long-tail navigation, while collaborative signals excel in high-density regions. 


\begin{table}[t!]
    \centering
    \caption{Complementarity Analysis: The Case of \textit{Real Steel}.}
    \label{tab:real_steel}
    \resizebox{\linewidth}{!}{
    \begin{tabular}{c|l|l}
        \toprule
        \textbf{Rank} & \textbf{Semantic View (Literal Trap)} & \textbf{Collaborative View (Latent Interest)} \\
        \midrule
        1 & Real Steel (Blu-ray) & Real Steel (Blu-ray) \\
        2 & Iron Man (Action) & \textbf{Pacific Rim (Mecha/Sci-Fi)} \\
        3 & Man of Steel (Superman) & Marvel's The Avengers \\
        4 & Heavy Metal (Sci-Fi) & Transformers: Dark of the Moon \\
        \bottomrule
    \end{tabular}
    }
\end{table}

\section{Related Work}

\noindent\textbf{The Status Quo: Paradigms of Semantic--Collaborative Integration.}
Integrating content semantics with collaborative signals is a long-standing theme in recommender systems. Early hybrid models such as CMF~\cite{CMF2008} and CTR~\cite{CTR2011} assumed a shared latent space governing both interactions and item text, naturally leading to content-to-collaborative mapping approaches like CB2CF~\cite{CB2CF2019}. In the LLM era, this goal has consolidated into two paradigms sharing a common premise: semantic and collaborative representations are globally bridgeable via low-complexity transformations. The \textit{Alignment Paradigm} enforces cross-view agreement via matching or contrastive objectives; RLMRec~\cite{RLMRec}, CoLLM~\cite{zhang2025collm}, and adapter-style frameworks~\cite{wang2025pad} explicitly encourage the semantic space to approximate collaborative geometry. The \textit{Homomorphism Paradigm} instead treats semantic representations as sufficiently expressive and maps them directly into recommendation vectors; ID-free systems such as UniSRec~\cite{hou2022towards} and ZESRec~\cite{ding2021zero} use text-derived item indices, while AlphaRec~\cite{AlphaRec} suggests collaborative signals are implicitly recoverable from LM representations via a low-capacity projector. Both paradigms either emphasize the shared intersection of views or implicitly require semantic substitution for collaboration, under-utilizing view-specific cues when overlap is limited.

\noindent\textbf{Complementarity in Multi-View Learning.}
Our perspective is grounded in multi-view learning theory, where different views contain both shared and private information. Classical frameworks such as CCA~\cite{CCA} and co-training~\cite{CoTraining1998} pursue agreement in a shared subspace rather than requiring full-space equivalence, and recent approaches explicitly disentangle shared from private factors to avoid view collapse~\cite{DMVAE,DSN}. These foundations suggest that forcing full geometric matching or assuming semantic recoverability of collaborative structure can be suboptimal when view-specific factors are substantial. We therefore advocate a fusion-centric stance that maintains view independence and combines evidence at the decision level. This treats fusion as both a practical baseline and a diagnostic: if simple fusion approaches an oracle upper bound, it provides direct evidence that preserving the full union of complementary private signals is more effective than forcibly collapsing views into a single geometry.

\section{Discussion and Conclusion}
\label{sec:discussion}

This paper questions the prevailing assumption that semantic and collaborative representations should be globally aligned through low-complexity transformations. We argue that this view is stronger than necessary and propose treating the two spaces as heterogeneous yet complementary views, each containing shared and private factors. Our empirical analyses confirm substantial complementarity, and even minimal fusion consistently outperforms alignment-based methods by exploiting the union rather than the intersection of signals.

Several challenges remain. The \textbf{sparsity-density trade-off} calls for adaptive mechanisms that modulate view contributions based on local data density. The reliance on \textbf{implicit union} rather than explicit shared-private disentanglement limits interpretability; techniques such as \textbf{VAE} or the \textbf{Information Bottleneck} could offer finer control. Finally, the \textbf{complexity dilemma} persists, as simple projections underfit while high-capacity mappings overfit, motivating richer geometric tools such as \textbf{optimal transport} or \textbf{Hyperbolic embeddings}. We hope this work encourages the community to move from enforcing alignment toward managing complementarity in multi-view recommendation.
\section*{Acknowledgements}
This work was supported by the National Science Fund for Excellent Young Scholars (Overseas) under grant No.\ KZ37117501, National Natural Science Foundation of China (No. \ 62306024), the Fundamental Research Funds for the Central Universities.
\balance
\bibliographystyle{ACM-Reference-Format}
{\bibliography{acmart}}


\end{document}